 \documentclass[aps,prl,twocolumn,superscriptaddress]{revtex4-1}
\usepackage{graphicx}
\usepackage{xcolor}
\usepackage{amsmath}
\usepackage[normalem]{ulem}

\begin{document}

% Use the \preprint command to place your local institutional report
% number in the upper righthand corner of the title page in preprint mode.
% Multiple \preprint commands are allowed.
% Use the 'preprintnumbers' class option to override journal defaults
% to display numbers if necessary
%\preprint{}

%Title of paper
\title{Interferometry for full temporal reconstruction of laser-plasma accelerator-based seeded free electron lasers}

% \author{M. Labat et al.}

\author{M. Labat}\email{marie.labat@synchrotron-soleil.fr}\affiliation{Synchrotron SOLEIL, L'Orme des Merisiers, 91 191 Gif-sur-Yvette, France}

\author{S. Bielawski}\affiliation{Univ. Lille, CNRS, UMR 8523 - PhLAM - Physique des Lasers Atomes et Molécules, F-59000 Lille, France.}
\author{A. Loulergue}\affiliation{Synchrotron SOLEIL, L'Orme des Merisiers, 91 191 Gif-sur-Yvette, France}
\author{S. Corde}\affiliation{Laboratoire d'Optique Appliqu\'ee, 181 chemin de la Huni\` ere, 91120 Palaiseau, France}
\author{M.E. Couprie}\affiliation{Synchrotron SOLEIL, L'Orme des Merisiers, 91 191 Gif-sur-Yvette, France}
\author{E. Roussel}\affiliation{Univ. Lille, CNRS, UMR 8523 - PhLAM - Physique des Lasers Atomes et Molécules, F-59000 Lille, France.}

%Collaboration name if desired (requires use of superscriptaddress
%option in \documentclass). \noaffiliation is required (may also be
%used with the \author command).
%\collaboration can be followed by \email, \homepage, \thanks as well.
%\collaboration{}
%\noaffiliation

\date{\today}

\begin{abstract}
The spectacular development of Laser-Plasma Accelerators (LPA) appears very promising for a free electron laser application. The handling of the inherent properties of those LPA beams already allowed controlled production of LPA-based spontaneous undulator radiation. Stepping further, we here unveil that the forthcoming LPA-based seeded FELs will present singular spatio-spectral distributions. Relying on numerical simulations and simple analytical models, we show how those interferometric patterns can be exploited to retrieve, in single-shot, the spectro-temporal content and source point properties of the FEL pulses.
\end{abstract}

% insert suggested keywords - APS authors don't need to do this
%\keywords{}

%\maketitle must follow title, authors, abstract, and keywords
\maketitle

%%%%%%%%%%%%%%%%%%%%%%%%%%%%%%%%%%%
% INTRODUCTION
%%%%%%%%%%%%%%%%%%%%%%%%%%%%%%%%%%%
Free Electron Lasers (FELs)~\cite{Madey_1971,Deacon_1977} deliver ultrashort, narrow-band and ultrabright pulses down to the hard X-ray range~\cite{Emma_2010,Ishikawa_2012,Weise_2017}, enabling breakthrough experiments in chemical, physical, and biological sciences. 
These light sources rely on relativistic electron beams wiggling in the periodic magnetic field of an undulator as gain medium. Interacting with the spontaneous radiation of the undulator or an external seed, the electrons experience an energy modulation at the resonance wavelength which is further transformed into a density modulation by dispersive elements. After this ``lethargy''~\cite{Bonifacio_1984} phase, the beam density modulation allows the emission of a coherent radiation which can then be exponentially amplified. A saturation is reached when the electrons energy loss is such that the resonance condition is violated, causing a red spectral shift of the FEL line~\cite{Wang_2009,Schneidmiller_2015}.

Chicanes as dispersive elements are used to speed up the electron beam energy to density modulation conversion~\cite{Csonka_1978,Vinokurov_1978}. Associated to electron beam energy chirping in the Radio-Frequency Accelerators (RFAs) structures producing those beams and then to frequency chirped seeds, they opened the door to FEL wavelength tunability~\cite{Shaftan-2005}, two-color operation~\cite{DeNinno_2013}, amplitude and phase control~\cite{Wu_2007,Lutman_2009}, spectro-temporal shaping~\cite{Gauthier_2015}, chirped-pulse compression~\cite{Wu_2007_b,Gauthier_2016} and temporal reconstruction~\cite{Deninno_2015}.

In high-gain FELs~\cite{Bonifacio_1984}, the initial radiation can be amplified by several orders of magnitudes up to saturation in one single pass.
The gain of those systems is mainly driven by the electron beam properties.
This is the reason why first FELs were developed using classical RFAs which presently deliver the requisite high-quality beams. However since those accelerators are typically several hundreds of meters long for a lasing in the X-ray range, the new Laser Plasma Accelerators (LPAs)~\cite{Tajima_1979} have been proposed as an alternative towards miniaturised FELs~\cite{Gruner_2007}. 

However from proposal to demonstration, the path is tortuous.
In LPAs, the intrinsic large divergence ($\approx$~few milliradians) and energy spread ($\approx$~few percent) of the electron beam~\cite{Couperus_2017,Tai_2018} dramatically limit the FEL gain. Several techniques can be implemented to reach the mandatory Pierce parameter $\rho_{fel}$~\cite{Bonifacio_1984} qualifying the FEL gain: use plasma lenses~\cite{Kuschel_2016,Thaury_2015,Tilborg_2015,Chiadroni_2018} and/or high gradient quadrupoles in the immediate vicinity of the LPA source to handle the large divergence~\cite{Eichner_2007,Mihara_2018,Ghaith_2018}, use a chicane~\cite{Couprie_2014,Maier_2012} or a transverse gradient undulator~\cite{Smith_1979,Huang_2012} to lower the slice energy spread, and/or use a chromatic matching~\cite{Loulergue_2015} to synchronize the beam maximum density with the radiation pulse propagation inside the undulator.
Few years after the first observation of LPA-based undulator radiation~\cite{Schlenvoigt_2008,Fuchs_2009,Lambert_2012,Anania_2014}, significative achievements in terms of electron beam transport and manipulation~\cite{Andre_NatCom_2018} turned a corner, allowing for control of the synchrotron radiation spectral properties~\cite{Ghaith_2019}.
LPA based FEL amplification is now within reach at least in the visible or ultra-violet range. To ease this next step, a seeded configuration should be considered, since it enables to speed up the lethargy phase, i.e. requires shorter undulator lengths and relaxes the transport issues.
Given the present LPAs performance, the first demonstration is more likely to be obtained in the low-gain~\cite{Elias_1976} regime as proposed in \cite{Liu_2019}. 

In this letter, we reveal that those forthcoming LPA based seeded FELs can present original spatio-spectral distributions that enable the full reconstruction of the temporal FEL pulse. Indeed, under simple conditions, the emitted radiation exhibits a clear red shift with respect to the input seed wavelength, leading to spectral interference fringes between this additional emitted radiation and the input seed. 
In the case of LPAs, the electron beam energy chirp, i.e. longitudinal sorting in energy, is not induced by the accelerating structure but by a dispersive transport of the high energy spread beam through a chicane. The resulting particles velocity gradient is so steep that the undulator dispersion drives the energy to density modulation conversion while simultaneously stretching the density modulation period (ahead travelling faster than tail particles). This effect is usually negligible on RFA based FELs, even though observed on a self-seeding configuration~\cite{Inoue_2019}. In the case of LPA beams however, the stretching is such that the final FEL wavelength can be red shifted by more than one seed linewidth. Due to the low-gain configuration, the red-shifted coherent emission is observed together with the seed, leading to strong intereferences in the spectral domain. As in \cite{Deninno_2015}, this interferometric pattern can then be used for temporal reconstruction of the FEL pulses.

\begin{figure}[htbp]
\includegraphics[width=\columnwidth]{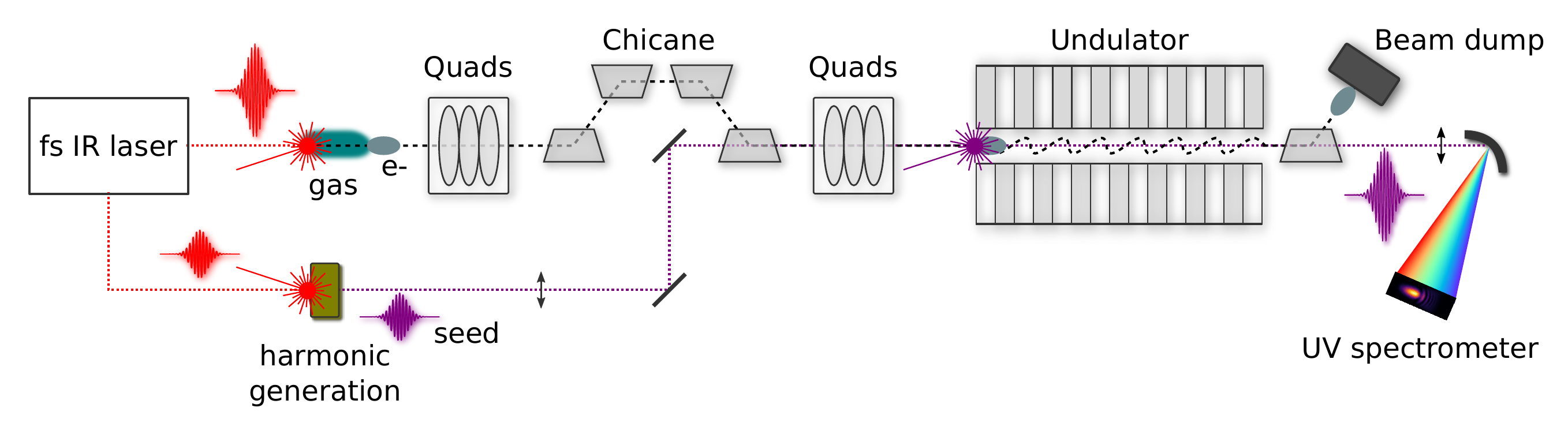}%
\caption{\label{fig:layout} Schematic of a LPA based seeded FEL. From one main femtosecond infrared laser unit, two pulses centred at 800~nm are delivered with independent controls. The main laser branch is focused in a gas medium for electron beam generation. The second branch is used for seed generation. At the undulator exit, the radiation is observed using an imaging spectrometer while the electron beam is dumped.}
\end{figure}
A typical setup for a LPA based seeded FEL is shown in Fig.~\ref{fig:layout}. In state-of-the-art LPAs, nonlinear plasma waves are driven in a gas by femtosecond pulses of a multi-TW laser system, allowing for the trapping and further acceleration of electron bunches. The properties of the generated electron beams strongly depend on the LPA implementation. We choose a set of parameters (see Table~\ref{tab:param}) compatible with the most recent achievements of the groups targeting both quality and reliability, mandatory for an FEL application~\cite{Couperus_2017,Weing-2012,Buck-2013,Wang-2012,Liu_2019}. 
With a short period undulator~\cite{chams_2017}, these typical 200~MeV beams can be used to target the vacuum ultra-violet range. We therefore consider a seed at 160~nm (see Table~\ref{tab:param}) generated in gas~\cite{Salieres_1995,Lambert_2008} using a branch of the main laser~\cite{Lambert_2008}. 
The electron beam is transported along a dedicated line allowing for a three steps manipulation~\cite{Loulergue_2015}: refocusing at the LPA exit using high gradient quadrupoles~\cite{Marteau_2017}, longitudinal stretching and energy chirping using a chicane and chromatic matching using standard quadrupoles. Locally driving the beam off-axis, the chicane is also used to insert a mirror for the seed injection on-axis.
Experimentally, an imaging spectrometer can provide the spatio-spectral distribution of the FEL radiation. Combined with adequate focusing optics, one can observe either the near- or far-field spectral distribution.
\begin{table}%[H] add [H] placement to break table across pages
\caption{\label{tab:param} Parameters for FEL simulation and modeling.}
\begin{ruledtabular}
\begin{tabular}{ccc}
& Parameter & Value \\
\hline 
Electron beam & Energy      & 200~MeV \\      
&Relative energy spread   & 4~$\%$   \\
&Charge $Q$          & 20~pC     \\
&Divergence $\sigma'_{x0,y0}$  & 1~mrad-rms         \\
%Size $\sigma_z$  	& 0.1~$\mu$m-FWHM       \\
&Emittance $\epsilon_{x0,y0}$  & 0.1~mm.mrad       \\
&Chicane dispersion $R_{56,c}$         & 2.5~mm       \\
\hline 
Undulator & Period $\lambda_u$ & 18~mm \\
& Deflection parameter $K$ & 1.864 \\
\hline
Seed laser & Central wavelength  & 160~nm \\ 
&Quadratic phase     & 1.15$\times 10^{-4}$~fs$^{-2}$ \\
&Duration            & 370~fs-FWHM \\
&Peak power          & 10~kW \\
&Waist position      & 0~m \\
&Rayleigh length     & 0.7~m \\
\end{tabular}
\end{ruledtabular}
\end{table}

\begin{figure}[h!]
\begin{center}
\includegraphics[width=\columnwidth]{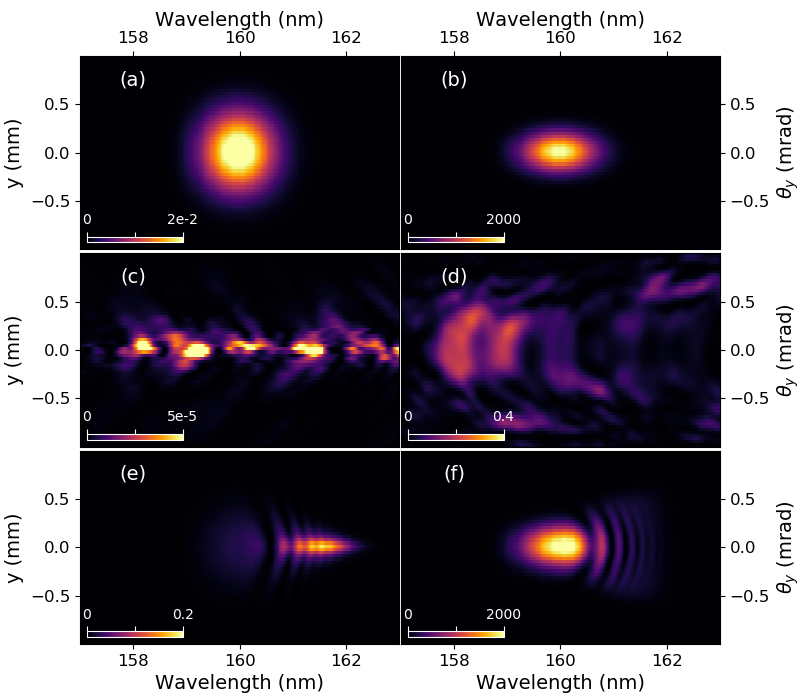}\\
\caption{\label{fig:2} Simulated spatio-spectral distribution of the radiation at the undulator exit in the near- (left) and far-field (right). (a,b) Seed only ($P_{seed}=10$~kW, $Q=0$~pC), (c,d) Spontaneous emission only ($P_{seed}=0$~W, $Q=20$~pC) and (e,f) FEL case ($P_{seed}=10$~kW, $Q=20$~pC). Other parameters from Table~\ref{tab:param}.}
\end{center}
\end{figure}

Using parameters of Table~\ref{tab:param}, a 6D electron beam is generated with a flat top profile in energy and Gaussian profiles in other dimensions. The distribution is then transported to the undulator entrance using second order transport formalism~\cite{Brown}. The FEL amplification is simulated with GENESIS~\cite{Reiche_1999} using this 6D electron beam distribution together with an amplitude and phase temporal description of the seed while the undulator magnetic field is defined by its period $\lambda_u$ and deflection parameter $K$ (see Table~\ref{tab:param}). The output radiation is then analyzed using Fourier optics (see supplemental materials) to predict the observable near- and far-field spatio-spectral distributions.
Typical simulation outputs are presented in Fig.~\ref{fig:2}. To identify the seed contribution, the charge $Q$ is first forced to zero. As expected, Gaussian distributions are obtained in the near- and far-fields (Fig.~\ref{fig:2}(a,b)). The seed power $P_{seed}$ is then forced to zero to distinguish the spontaneous emission contribution (Fig.~\ref{fig:2}(c,d)) expected at a resonance wavelength $\lambda_R=\lambda_u/(2\gamma^2)(1+K^2/2+\theta^2\gamma^2)$ for a mono-energetic beam with a Lorentz factor $\gamma$,  $\theta$ being the observation angle. Hence the observed broad band emission results from the large energy spread (different $\gamma$) while the ``moon shapes'' in the far-field result from the off-axis contribution in $\theta^2$~\cite{Giannessi_2013}. When the seed overlaps the electron beam, i.e. in the FEL case, a significant red shift ($\approx$1.5~nm) appears (Fig.~\ref{fig:2}(e,f)), together with high contrast fringes.

The standard FEL red spectral shift due to beam energy loss can only be responsible here for sub-nanometer spectral shifts (0.002~nm in the case of Table~\ref{tab:param}). 
In order to understand the process underlying here, we reconsider the analytical treatment of the electron beam longitudinal motion along the undulator. For a relativistic electron of energy $\gamma mc^{2}$ travelling in a purely vertical undulator magnetic field, the Lorentz equation in the horizontal direction can be integrated to derive the average longitudinal velocity~\cite{Huang_2007}:
\begin{equation}
\tilde{v}_z \approx c \left[ 1 - \frac{1}{2\gamma^2}\left( 1 + \frac{K^2}{2}\right) \right]
\label{eq:vz_avg}
\end{equation}
The time for an electron to travel the distance $z$ in the undulator is simply $t = z/\tilde{v}_z$ so that the difference of travel time $dt$ for electrons of different energies $d\gamma$ close to $\gamma$ is:
\begin{equation}
dt = z \frac{d}{d\gamma}\left(\frac{1}{\tilde{v}_z}\right) d\gamma
\approx - \frac{z}{c}\left(1+\frac{K^{2}}{2}\right) \frac{d\gamma}{\gamma^{3}}
\end{equation}
Thus, electrons with different energies induce a stretching $\Delta \lambda$ of the beam due to the longitudinal dispersion $R_{56,u}(z)$ of the undulator:
\begin{equation}
\Delta \lambda = c \Delta t = - z \left(1+\frac{K^{2}}{2}\right) \frac{1}{\gamma^{2}} \frac{\Delta\gamma}{\gamma} = R_{56,u}(z) \frac{\Delta\gamma}{\gamma}
\label{eq:red_shift1}
\end{equation}
In a first approximation, the longitudinal energy sorting of the electrons introduced in the chicane is given by:
\begin{equation}
\Delta s = R_{56,c} \frac{\Delta\gamma}{\gamma}
\label{eq:r56_chic}
\end{equation}
with $R_{56,c}$ the chicane strength or dispersion.
Considering now a beam portion of one seed laser period, i.e. $\Delta s = \lambda_{seed}$, and inserting Eq.~(\ref{eq:r56_chic}) into Eq.~(\ref{eq:red_shift1}), we find that this beam portion is increased by:
\begin{equation}
\label{eq:delta_2}
\begin{split}
\Delta \lambda (z) & = \frac{R_{56,u}(z)}{R_{56,c}} \lambda_{seed} 
	 = - \frac{z\left(1+\frac{K^{2}}{2}\right)}{\gamma_{0}^{2}R_{56,c}}  \lambda_{seed}
\end{split}
\end{equation}
The initial $\lambda_{seed}$ density modulation is stretched proportionnally to the undulator and chicane dispersion ratio up to $\lambda_{seed} + \Delta \lambda(z)$. Since the coherent radiation is emitted at the final density modulation period, the FEL central wavelength is found shifted by $\Delta \lambda$. Because the density modulation is not present immediately at the undulator entrance, i.e. at $z=0$, this model can be further refined replacing $z$ in Eq.~(\ref{eq:delta_2}) by $z-z_i$, where $z_i$ is the ``effective'' starting point of the red shifting.
Systematic analysis of the simulations reveals that $z_i$ could be defined as the location for which the coherent emission intensity becomes $\approx 0.5$ times the initial seed intensity.

\begin{figure}[h!]
\includegraphics[width=\columnwidth]{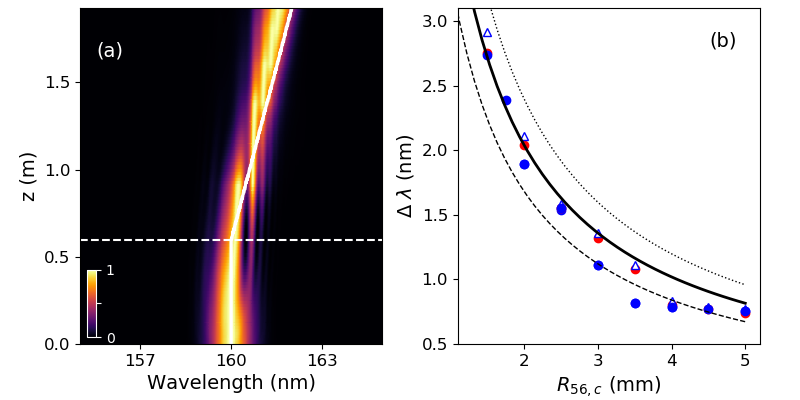}\\
\caption{\label{fig:3} (a) Normalized simulated spectrum along the undulator for a charge of 50~pC, with theoretical prediction of the peak wavelength evolution (solid white line) using a starting point $z_i$ (dashed line) where the FEL intensity reaches 0.5 times the seed intensity. (b) Simulated spectral shift $\Delta \lambda$ versus the chicane dispersion $R_{56,c}$ for $Q=20$~pC and $\sigma'_{x0,y0}=1$~mrad (blue dot), $Q=40$~pC and $\sigma'_{x0,y0}=1$~mrad (red dot), $Q=20$~pC and $\sigma'_{x0,y0}=0.5$~mrad (blue triangle). Analytical prediction of the spectral shift for various starting points: $z_{i}=0.25$~m (dashed line), $z_{i}=0.5$~m (solid line), $z_{i}=0.75$~m (dotted line).
$R_{56,c}$ set to $1.5$~mm corresponding to a value compatible with the insertion of a seed mirror in the chicane.}
\end{figure}
Fig.~\ref{fig:3}(a) shows a typical evolution of the FEL spectrum along the undulator. The FEL wavelength drifts away from the initial seed wavelength, following the analytical prediction of Eq.~(\ref{eq:delta_2}) (solid line). The horizontal dashed line corresponds to the starting point $z_{i}$ in the model. Figure~\ref{fig:3}(b) summarises the FEL central wavelengths obtained in simulation as a function of the chicane strength for various electron beam charges and divergences. The final wavelengths are found well within the theoretical predictions.

\begin{figure}[h!]
\begin{center}
\includegraphics[width=\columnwidth]{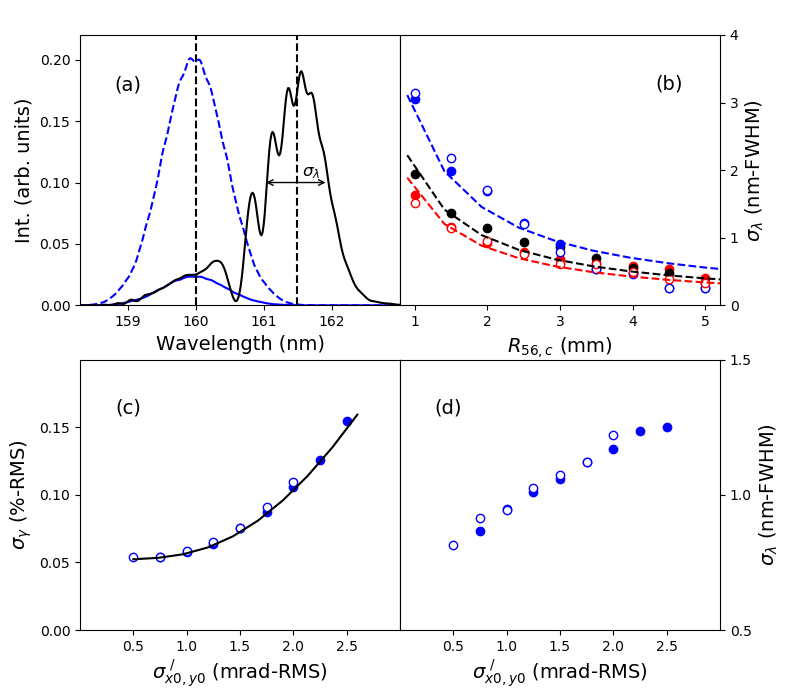}\\
\caption{\label{fig:4} (a) Simulated on-axis spectrum of the seed at undulator entrance (dashed blue line) and exit (continuous blue line), and of the FEL at undulator exit (continuous black line). Dash lines: seed and FEL central wavelengths (maximum intensity location). (b) Simulated FEL linewidth versus chicane dispersion for $\sigma'_{x0,y0}$=0.5~mrad-rms (blue dots), $\sigma'_{x0,y0}$=1.0~mrad-rms (black dots), $\sigma'_{x0,y0}$=2.0~mrad-rms (red dots), $\epsilon_{x0,y0}$=0.1~mm.mrad (full dots) and $\epsilon_{x0,y0}$=0.5~mm.mrad (empty dots), using $Q$=40~pC. Fits (dashed lines) of the linewidth with $f(a)=a/R_{56,c}$ where $a$=1.7, 2.0 and 2.8 for respectively the 0.5, 1.0 and 2.0~mrad cases.
(c) Simulated electron beam slice energy spread inside the undulator versus beam initial divergence using same markers as for (b) at $R_{56,c}$=2.5~mm and calculated using Eq.~\ref{eq:espread} (black line) with $R_{522}=R_{544}$=-0.23~m. (d) Simulated FEL linewidth versus beam initial divergence using same markers as for (b) at $R_{56,c}$=2.5~mm. Other parameters from Table~\ref{tab:param}.}
\end{center}
\end{figure}
The analysis of the FEL spectral content can then be pushed further. The typical on-axis radiation spectra at the undulator entrance, i.e. of the seed (blue dashed line), and undulator exit, i.e. of the FEL (continuous black line), are shown in Fig.~\ref{fig:4}(a). In addition to the red shift amplitude, the FEL linewidth $\sigma_{\lambda}$ can be analysed as a function of the chicane dispersion (Fig.~\ref{fig:4}(b)). It clearly comes out that, for a given charge, the FEL linewidth can be controlled by the chicane dispersion scaling as $1/R_{56,c}$. 
The FEL linewidth also appears strongly correlated to the initial beam divergence (color trends) while hardly sensitive to the beam emittance (full versus empty dots). In the low gain regime, according to the resonance wavelength expression, a relative energy spread of $\sigma_{\gamma}$ can lead to an inhomogeneous broadening of the relative linewidth by 2$\sigma_{\gamma}$. 
In the case of LPA beams, using simple analytical formula~\cite{Labat_prstab_2018}, the electron beam slice energy spread at the undulator entrance can be approximated by:
\begin{equation}
\label{eq:espread}
\sigma_{\gamma} \approx \frac{1}{R_{56,c}} \sqrt{\sigma_{z0}^2 + 2R_{522}^2\sigma_{x0}^{'4}+2R_{544}^2\sigma_{y0}^{'4}}
\end{equation}
with $R_{522,544}$ the transport matrix coefficicents, $\sigma_{z0}$ the length and $\sigma_{x0,y0}'$ the divergence of the beam at the LPA source.
Hence, after transport, the initial divergence turns out to be the dominant contribution to the electron beam slice energy spread. The slice energy spread inferred from the electron beam distribution at the undulator exit follows in satisfying agreement with this prediction (Fig.~\ref{fig:4}(c)). Such a correlation of the FEL linewidth with the initial beam divergence (Fig.~\ref{fig:4}(d)) can be seen as a special feature of the LPA based FELs. 

\begin{figure}[h!]
\begin{center}
\includegraphics[width=\columnwidth]{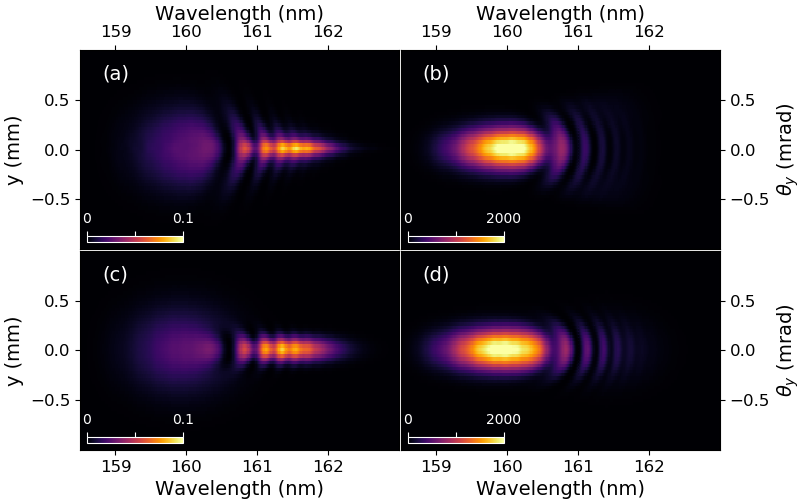}\\
\includegraphics[width=\columnwidth]{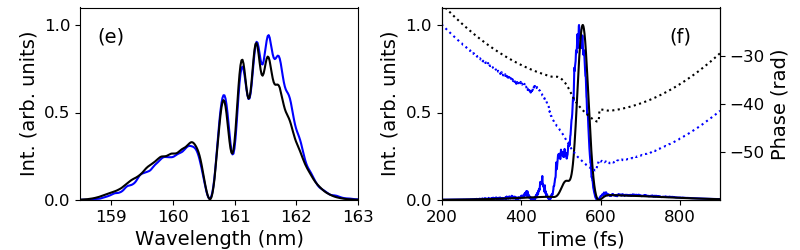}\\
\caption{\label{fig:5} Near- (a,c) and far-field (b,d) from (top) simulations and (bottom) analytical model fitting simulations with 
$E_{0,coh}/E_{0,seed}=2.35$, $\sigma_{t,coh}=17.3$~fs, $\phi_{coh}=-9.3 \times 10^{6}$, $\omega_{0,coh}=2\pi c/161.5$~nm, $C_{1,coh}=3.6\times 10^{-4}$~fs$^{-2}$, waist position $0.23$~m before undulator exit, $z_{R,coh}=0.25$~m and $\tau$=50~fs (see supplemental material). Simulated (blue) and fitted (black) on-axis FEL spectral intensity (e) and field intensity and phase (f). Other parameters from Table~\ref{tab:param} with $Q$=15~pC.}
\end{center}
\end{figure}

The on-axis FEL spectra (Figs.~\ref{fig:3}(a,b)-~\ref{fig:4}(a)) present fringes that are also visible in the 2D spatio-spectral distribution (Fig.~\ref{fig:2}(e,f)). 
Most features of this interferometric pattern can be simply understood modeling the exit radiation as a superposition of two independent Gaussian light pulses: the seed laser ($E_{seed}$) and a coherent contribution ($E_{coh}$). Each light pulse electric field is described in 3D as the product of a transverse field $E_\text{trans}(x,y,z)$ and a longitudinal field $E_\text{temp}(t)$, where ($x$,$y$) stand for the coordinates in the transverse plane perpendicular to the propagation axis $z$ and $t$ refers to the time (see supplemental material). The longitudinal electric field can be expressed as:%
\begin{equation}
	E_\text{temp}(t) = e^{-t^2/(4\sigma_t^2)} e^{i (\phi + \omega_0 t + C_1 t^2)}
	\label{eq:Ez}
\end{equation}
where $\sigma_t$ is the RMS pulse duration, $\phi$ a phase offset, $\omega_{0}$ the optical carrier frequency and $C_{1}$ the quadratic phase.
Assuming that the FEL results from the superposition of those two fields, its electric field is finally obtained according to:
\begin{equation}
E_{fel} (x, y, z, t) = E_{seed}(x, y, z, t) + E_{coh}(x, y, z+ \Delta z, t-\tau)
\label{eq:Efel}
\end{equation}
with $\Delta z$ the difference between seed and coherent component waist positions and $\tau$ their relative delay. The ratio $E_{coh}/E_{seed}$ corresponds to the FEL ``gain''.
Using Fourier optics, calculation of both near- and far-field spectral distributions is then straightforward (see supplemental material). 
Assuming that the seed properties are known and using the double pulse model (Eq.~(\ref{eq:Efel})), the simulated near-field pattern (Fig.~\ref{fig:5}(a,b)) is fitted using a standard least mean square method. The result of the fit is shown in Fig.~\ref{fig:5}(c,d). Both near- and far-field distributions present a good agreement with the initial simulations. The few percent fit accuracy is illustrated in Fig.~\ref{fig:5}(e), comparing the near-field on-axis spectra (i.e. at $y=0$~mm).
Among the fitted parameters are found key features such as the FEL source point and Rayleigh length (respectively 0.23~m before the undulator exit and 0.25~m in the case of Fig.~\ref{fig:5}(c-d)). Finally, the reconstructed FEL intensity and phase in the time-domain are displayed in Fig.~\ref{fig:5}(f), again in very good agreement with the initial simulation. 

In conclusion, we presented simulation results of a feasible LPA based seeded FEL. 
We showed that it presents original spatio-spectral distributions, with red shifted interference fringes. Those features were explained using simple considerations, allowing a deep understanding of the physical effects underlying.
Moreover, we demonstrated that our double pulse model can provide with key features on the FEL source point as well as the full temporal reconstruction of the FEL pulse on a single-shot basis. Such tools should ease the path towards the very first LPA based FEL demonstration. 
We verified that this work can be extended to FELs in the visible or XUV range. Fringes can be indeed obtained as long as the red shift amplitude is of the order of the seed linewidth, and as the gain remains below two orders of magnitude.

\begin{acknowledgments}
This work was partially supported by the European Research Council for the Advanced Grants COXINEL (340015, PI: M.-E. Couprie), the LABEX CEMPI (ANR-11-LABX-0007), Ministry of Higher Education and Research, Hauts de France council and European Regional Development Fund (Contrat de Projets Etat-Region CPER Photonics for Society P4S). The authors acknowledge the SOLEIL team and the LPA team of Laboratoire d'Optique Appliqu\'ee for the collaboration on COXINEL.  
\end{acknowledgments}

\bibliography{bib_file}

\end{document}